\documentclass[10pt,a4paper,twoside,leqno]{bull}
\usepackage{amsfonts,amssymb}
\usepackage{graphicx}
\usepackage{cite}
\usepackage[T1]{fontenc}

 4
\font\ebf=cmbx8

\usepackage{graphicx}

\begin{document}

\renewcommand{\thepage}{[\arabic{page}]}



\newcounter{theo}
\newtheorem{theor}[theo]{Theorem}
\newcounter{lem}
\newtheorem{lemma}[lem]{Lemma}
\newtheorem{theorem}{Theorem}[section]
\newtheorem{corollary}[theorem]{Corollary}
\newtheorem{definition}[theorem]{Definition}
\newtheorem{remark}[theorem]{Remark}
\def\nine{\small}
\def\b{\break}

\thispagestyle{empty}

{\baselineskip=15pt

\begin{centering}
{\Large\bf  Erroneous solution of
three-dimensional (3D) simple orthorhombic Ising lattices}

\vspace{0.3cm}



{\Large\bf  Erroneous solution of three-dimensional (3D) simple
orthorhombic Ising lattices II
}

\vspace{0.3cm}



\noindent {\Large\bf  Erroneous solution of three-dimensional (3D) simple
orthorhombic Ising lattices III
}

\vspace{0.5cm}

 {\Large\it Jacques H.H. Perk}

\vspace{0.5cm}

{\large Department of Physics, Oklahoma State University,\\
145 Physical Sciences, Stillwater, OK 74078-3072, USA\\
E-mail: perk@okstate.edu}

\vspace{0.5cm}

{\large\it Invited contribution to the proceedings of}

\vspace{0.3cm}

{\large Hypercomplex Seminar 2012: (Hyper)Complex Function Theory,\\
Regression, (Crystal) Lattices, Fractals, Chaos, and Physics}

\vspace{0.3cm}

{\large Mathematical Conference Center at B\k{e}dlewo (Poland), July 8--15, 2012}

\vspace{0.3cm}

{\large\it published in}

\vspace{0.3cm}

{\large Bulletin de la Soci\'et\'e des Sciences et des Lettres de \L\'od\'z\\
S\'erie: Recherches sur les D\'eformations \textbf{62} no.\ 3 (2012) 45--59, 71--74}

\end{centering}

\vspace{0.5cm}

\noindent {\large\bf Alternative abstract}

\vspace{0.3cm}

{\large The first paper is an invited comment on arXiv:1110.5527
presented at Hypercomplex Seminar 2012 and on sixteen earlier published
papers by Zhidong Zhang and Norman H.~March. All these works derive
from an erroneous solution of the three-dimensional Ising model published
in 2007. A self-contained detailed rigorous proof  is presented that the final
expressions in this work are wrong and that the conjectures on which they
are based consequently fail. Further errors and shortcomings in the follow-up
works are also pointed out.

The second paper is a comment on the response arXiv:1209.3247  by Zhang and March.
The third paper gives further comments.}

}
\newpage



\noindent {\it Jacques H.H. Perk}

\vspace{0.5cm}

\noindent {\bf  Erroneous solution of
three-dimensional (3D) simple orthorhombic Ising lattices}

\vspace{0.4cm}

\noindent {\ebf Summary}

{\small Thirteen follow-up papers by Zhang and March perpetuate the errors
of a 2007 paper by Zhang, which was based on an incorrect application
of the Jordan--Wigner transformation and presents final expressions
that contradict rigorously established exact results. The presentation
given here can be used as a brief mathematical introduction to the
Ising model for nonexperts.}


\section{Introduction}
In a very long paper \cite{zdz} published in 2007 Z.-D. Zhang claims to
present the exact solution of the free energy per site and of the
spontaneous magnetization of the three-dimensional Ising model in
the thermodynamic limit. This claim has been shown to be false
\cite{wmfc,zdz2,wmfc2,jhhp,zdz3,jhhp2} and we shall show here that
very little original work, if any, in \cite{zdz} can be salvaged.

The principal reason why the outcomes of \cite{zdz} are wrong is that
they contradict exactly known series expansion results \cite{wmfc,jhhp}.
Several references were cited in \cite{wmfc,jhhp} which show that \cite{zdz}
violates rigorously established theorems. As these cited theorems are formulated
for very general lattice models with rather general interactions, requiring
complicated notations and such concepts as Banach spaces and Banach algebras,
it takes some effort to check that every needed detail is there to make
the proof rigorous.

Therefore, we present here a simpler self-contained presentation, restricted
to the three-dimensional Ising model on a simple cubic lattice, which can
be used as a short introduction for nonexperts interested in this model.

\begin{definition}\rm
The isotropic Ising model on $\mathbb{Z}_n^{\,3}$,
a periodic $n\times n\times n$ lattice with $N=n^3$ sites
$i=(i_x,i_y,i_z)$ on a 3-torus,
is defined by its configuration space
\begin{equation}
\mathbb{T}^3\supset\mathbb{Z}_n^{\,3}\rightarrow\{\pm1\}^N,\qquad
i\mapsto\sigma_i=\pm1,\mbox{ for }i\in\mathbb{Z}_n^{\,3},
\end{equation}
and its interaction energy
\begin{equation}
\mathcal{H}_N:\mathbb{Z}_n^{\,3}\rightarrow\mathbb{C},\qquad
\mathcal{H}_N=\mathcal{H}_N(\{\sigma_i\})=
-J\sum_{\langle i,j\rangle}\sigma_i\sigma_j
-B\sum_{i}\sigma_i,
\label{ham}\end{equation}
where the sum over $\langle i,j\rangle$ is over all nearest-neighbor
pairs of sites $i$ and $j$, $J$ is the interaction strength
and $B$ is the scaled magnetic field. Sites $i$ and $j$ are
nearest-neighbor (nn) sites, if and only if
\begin{equation}
(i_x-j_x,i_y-j_y,i_z-j_z)=(\pm1,0,0),\;(0,\pm1,0),\mbox{ or }(0,0,\pm1)
\mbox{ mod }n.
\end{equation}
\end{definition}
\begin{remark}\rm
The generalization to the orthorhombic lattice is straightforward,\b
replacing $n$ by $n,n',n''$ and $J$ by $J,J',J''$ for the three
lattice directions. We\b consider the isotropic lattice for the sake
of simplicity of arguments, as this\b special case suffices to disprove
Zhang's claims \cite{zdz}.
\end{remark}

\begin{definition}\rm
Given a function $A\equiv A(\{\sigma\})$ of
the spin configuration, its\b expectation value is
\begin{equation}
\langle A\rangle_N^{\phantom{y}}=\frac1{Z_N}
\sum_{\{\sigma_i=\pm1\}}A\,\mathrm{e}^{-\beta\mathcal{H}_N},\qquad
\langle A\rangle=\lim_{N\to\infty}\langle A\rangle_N^{\phantom{y}},
\end{equation}
where the partition function,
\begin{equation}
Z_N=\sum_{\{\sigma_i=\pm1\}}\mathrm{e}^{-\beta\mathcal{H}_N},
\label{Z_N}\end{equation}
is a state sum taken over all $2^N$ spin configurations, while
$\beta=(kT)^{-1}$ with $T$ the absolute temperature and
$k$ Boltzmann's constant. If $\beta$, $J$, and $B$ are real,
then $\rho(\{\sigma\})=\mathrm{e}^{-\beta\mathcal{H}_N}/Z_N$ is
the Boltzmann-Gibbs canonical probability distribution.
\end{definition}

\begin{definition}\rm
The free energy per site $f_N$ and its infinite system limit $f$ are given by
\begin{equation}
-\beta f_N=\frac1N\log Z_N,\qquad f=\lim_{N\to\infty}f_N,
\label{f_N}\end{equation}
whereas the spontaneous magnetization is defined by
\begin{equation}
I=\lim_{B\downarrow0}\lim_{N\to\infty}\langle\sigma_{i_0}\rangle_N^{\phantom{y}}
=\lim_{B\downarrow0}\lim_{N\to\infty}\frac1{Z_N}
\sum_{\{\sigma_i=\pm1\}}\sigma_{i_0}\,\mathrm{e}^{-\beta\mathcal{H}_N},
\label{magn}\end{equation}
with $i_0$ any of the $N$ lattice sites, as the lattice is chosen periodic.
The pair-correlation function of spins at sites $i$ and $j$ is
$\langle\sigma_i\sigma_j\rangle$,
\end{definition}
\begin{remark}\rm
As in \cite{zdz}, we shall concentrate on the zero-field ($B=0$)
thermodynamic limit ($\lim_{N\to\infty}$). The order of limits in
(\ref{magn}) was used implicitly in Yang's paper \cite{Yang} on
the spontaneous magnetization of the square-lattice Ising model
\cite{Onsager,Kaufman,KO}.\b With the opposite order of limits the
result is identically zero. An alternative\b definition is
$I^2=\lim\langle\sigma_i\sigma_j\rangle|_{B=0}^{\phantom{y}}$ in
the limit of infinite separation of sites $i$ and $j$ \cite{MPW}.
\end{remark}

In \cite{zdz} Zhang starts out mimicking the treatment of the
two-dimensional Ising model by Onsager and Kaufman \cite{Onsager,Kaufman,KO},
in order to calculate the free energy,\b magnetization and pair correlation
of the three-dimensional case. Even though Zhang made two early errors
in \cite{zdz}, while transforming to Clifford algebra operators
and treating boundary terms \cite{jhhp}, he claims \cite{zdz3} that these
are overcome by two\b conjectures. But these conjectures are based on no
serious evidence whatsoever and the resulting expressions for the free
energy and magnetization \cite{zdz} are demonstrably incorrect, as they
fail the series test \cite{wmfc,jhhp}.

First, in section 2, a detailed account will be given of the rigorous
results of the 1960s violated by Zhang's work. Theorems \ref{corrtheorem}
and \ref{FreeEn} provide rigorous proof of the correctness of the series
test. Then, in section 3, further comments will be presented, including
several on the follow-up work by March and Zhang \cite{mz1,mz2,mz3,%
mz4,mz5,mz6,mz7,mz8,mz9,mz10,mz11,mz12,mz13}, which contain several
additional errors and misleading statements.

\section{Some rigorous results of the 1960s revisited}

In recognizing the criticisms to which his work in \cite{zdz}
has been subjected in\b \cite{wmfc,wmfc2,jhhp,jhhp2}, Zhang (supported
more recently by Norman H.~March) has argued that the usual
high-temperature series expansions \cite{AF}, renormalization group
treatments \cite{epsilon,PV}, and Monte Carlo simulations \cite{fb,ggkr},
fail to apply in the vicinity of infinite temperature owing to
singular behavior and Yang--Lee zeros \cite{YL,LY}
present even in the thermodynamic limit. Hence, it is argued,
such criticisms are not applicable as a basis for criticizing
the quite different conclusions he has reached. See specifically
the claims Zhang has made in the second paragraph of \cite{zdz2}, 
and in the second half of page 766 of \cite{zdz3}, as well
in section 5 of \cite{mz4}, second half of page 534.
The aim of this section is to show specifically by a detailed
mathematical analysis that there is no credibility at all in these claims.

In fact, five decades ago several theorems were published and
supported by fully rigorous proofs that underpin the validity
of the criticisms of Zhang's work, see, e.g., \cite{Ruelle,Griffiths}
for review. Nevertheless, let us here take the reader through
a simplified treatment especially tailored to apply to the point
at issue, namely, the statistical mechanics of the Ising model on
a cubic lattice with periodic boundary conditions.

\subsection{Free energy per site of a finite system vs.\ its large-system limit }

The proof of the thermodynamic (infinite system-size) limit of
the free energy typically uses the following lemma, see e.g.\ (2.15)
in \cite{Ruelle}:
\begin{lemma}
\begin{equation}
|\log\mathrm{Tr}\,\mathrm{e}^A-\log\mathrm{Tr}\,\mathrm{e}^B|
\le \|A-B\|,\quad\mbox{for $A$ and $B$ hermitian.}
\label{bogoliubov}
\end{equation}
\label{bogol}
\end{lemma}

\Proof
The proof follows immediately working out
$$\log\mathrm{Tr}\,\mathrm{e}^A-\log\mathrm{Tr}\,\mathrm{e}^B=
\int_0^1\frac{d}{dh}\log\mathrm{Tr}\,\mathrm{e}^{B+h(A-B)}dh=
\int_0^1\frac{\mathrm{Tr}\,(A-B)\mathrm{e}^{B+h(A-B)}}
{\mathrm{Tr}\,\mathrm{e}^{B+h(A-B)}}dh,$$
where the last integrand is an expectation value. (In this paper
we only need to consider the commuting case that $A$ and $B$
are diagonal matrices.)
\endproof

\begin{theor}
The free energy per site $f_N$ converges uniformly to a limit $f$
as the system size becomes infinite
for $\beta J$ real and bounded.
\label{thermlim}\end{theor}

\Proof
In order to prove this we must estimate $|f_N-f_M|$ for $N,M>N_0$,
with $N_0$ sufficiently large. Here we do that only for
periodic cubic lattices $N=n^3$, $M=m^3$ and compare with
the larger periodic cubic lattice of size $NM=(nm)^3$. By
changing a subset of the interactions we can change the larger
lattice into $N$ identical cubes of the size $M$ lattice, or
the other way around. The proof is then provided by counting
the changed interactions and by using Lemma \ref{bogoliubov}.
In our case, the trace in the lemma is just the sum over spin
configurations and the norm the maximum over all configurations.
\endproof

\begin{remark}\rm
Lemma~\ref{bogol} can also be used to show that the free energy $f$
does not depend on boundary conditions in the large system limit
with different shapes than cubes, provided it is taken in the sense
of van Hove, see e.g.\ \cite{Ruelle,Griffiths} for details.
\end{remark}

\begin{remark}\rm
The proof of Theorem \ref{thermlim} gives a rigorous bound on the
difference of the free energy per site of a finite system and its
large-system limit. It can therefore be used to estimate the
accuracy of finite-size calculations using e.g.\ Monte Carlo simulations.
\end{remark}

\subsection{Analyticity of the correlation functions and their thermodynamic limits}

\begin{lemma}
The partition function $Z_N$ (\ref{Z_N}) is a Laurent polynomial
in $\mathrm{e}^{\beta J}$, so that $\beta f_N$ is singular only
for the zeros of this Laurent polynomial and for
$\mathrm{e}^{\beta J}=\infty$. As $Z_N$ is a sum of positive
terms for real $\beta J$, it cannot have zeros on the real axis.
\end{lemma}

We will show that the zero
closest to $\beta J=0$ (or $\mathrm{e}^{\beta J}=1$) in the
complex $\beta J$ plane is uniformly bounded away, i.e.\
$Z_N\ne0$ for all $|\beta J|<K_0$ and all $N$ for some fixed
$K_0$. This means that $f_N$ can be expanded in a power series
in $\beta J$ that is absolutely convergent for $|\beta J|<K_0$
and uniform in $N$. It is well known that more and more
coefficients become independent of $N$ as $N$ increases.
Together this implies that the limiting $f$ also has a
power series in $\beta J$ with radius of convergence at
least $K_0$.

We continue by deriving a lower estimate for $K_0$.
Most proofs of the analyticity of free energies and
correlation functions use linear correlation identities
of Schwinger--Dyson type, known under such names as the
BBGKY hierarchy, Mayer--Montroll or Kirkwood--Salzburg
equations. We could use \cite{lp} and \cite{gmr}. But
instead, let me give an alternative proof using an identity of
Suzuki \cite{suz1,suz2}, restricted to the isotropic
Ising model on a simple cubic lattice with periodic boundary
conditions and of arbitrary size, as this method also can be
used to generate the coefficients of the high-temperature series.
More precisely, using the canonical definition of the
expectation value of a function $A\equiv A(\{\sigma\})$ of
the spin configuration,
we have the correlation identity \cite{suz1,suz2}:

\begin{lemma} (M. Suzuki, 1965 \cite{suz1,suz2})
\begin{equation}
\bigg\langle\prod_{i=1}^m\sigma_{j_i}\bigg\rangle_N=
\frac1m\sum_{k=1}^m
\bigg\langle\bigg(\prod_{i=1\atop i\ne k}^m\sigma_{j_i}\bigg)
\tanh\bigg(\beta J\!\sum_{l\;{\rm nn}\;j_k}\!
\sigma_{l}\bigg)\bigg\rangle_N,
\label{suzid}\end{equation}
where $j_1,\ldots,j_m$ are the labels of $m$ spins and $l$
runs through the labels of the six spins that are nearest
neighbors of $\sigma_{j_k}$.
\end{lemma}

\Proof
The proof of (\ref{suzid})
is easy summing over spin $\sigma_{j_k}$ in the numerator
of the expectation value, i.e.,
\begin{equation}
\sum_{\sigma_{j_k}=\pm1}\sigma_{j_k}
\mathrm{e}^{\beta J\sum_{l\;{\rm nn}\;j_k}\sigma_{j_k}\sigma_l}=
\tanh\bigg(\beta J\!\sum_{l\;{\rm nn}\;j_k}\!\sigma_l\bigg)
\sum_{\sigma_{j_k}=\pm1}
\mathrm{e}^{\beta J\sum_{l\;{\rm nn}\;j_k}\sigma_{j_k}\sigma_l}.
\end{equation}
Averaging over $k$ has been added
in (\ref{suzid}), so that all spins are treated equally,
consistent with the periodic boundary conditions. The lemma is also valid
without that.
\endproof

Next we use
\begin{lemma}
\begin{equation}
\tanh\bigg(\beta J\sum_{l=1}^6\sigma_l\bigg)=
a_1\sum_{(6)}\sigma_l
+a_3\sum_{(20)}\sigma_{l_1}\sigma_{l_2}\sigma_{l_3}+a_5
\sum_{(6)}\sigma_{l_1}\sigma_{l_2}\sigma_{l_3}\sigma_{l_4}\sigma_{l_5},
\label{suzidt}\end{equation}
where the sums are over the 6, 20, or 6 choices of choosing 1, 3, or 5
spins from the given $\sigma_1,\ldots,\sigma_6$. It is easy to check that
the coefficients $a_i$ are
\begin{eqnarray}
&&a_1=\frac{t(1+16t^2+46t^4+16t^6+t^8)}{(1+t^2)(1+6t^2+t^4)(1+14t^2+t^4)},
\qquad a_3=\frac{-2t^3}{(1+t^2)(1+14t^2+t^4)},\nonumber\\
&&a_5=\frac{16t^5}{(1+t^2)(1+6t^2+t^4)(1+14t^2+t^4)},
\qquad t\equiv\tanh(\beta J).
\label{suzida}\end{eqnarray}
The poles of the $a_i$ are at $t=\pm{\rm i}$, $t=\pm(\sqrt{2}\pm1){\rm i}$,
and $t=\pm(\sqrt{3}\pm2){\rm i}$. It can also be verified, e.g. expanding
the $a_i$ in partial fractions, that the series expansions of the $a_i$ in
terms of the odd powers of $t$ alternate in sign and converge absolutely
as long as $|\beta J|<\arctan(2-\sqrt{3})=\pi/12$.
\end{lemma}

\Proof
Clearly, the tanh in (\ref{suzidt}) can be expanded as done. Replacing all
six spins, $\sigma_l$ by $-\sigma_l$, shows that no terms with an even
number of spins occur. Also,\b permutation symmetry allows only three
different coefficients. Multiplying (\ref{suzidt}) with one, three, or
five spins $\sigma_l$ and then summing over all $2^6=64$ spin states,
is one way to derive (\ref{suzida}). It is then straightforward to verify
the following partial fraction expansions,
\begin{eqnarray}
&&a_{1,5}=\frac1{24}\Big(\frac{p_1t}
{1+(p_1t)^2}+\frac{p_2t}{1+(p_2t)^2}\Big)
\pm\frac{\sqrt{2}}{8}\Big(\frac{p_3t}
{1+(p_3t)^2}+\frac{p_4t}{1+(p_4t)^2}\Big)\nonumber\\
&&\hspace{2cm}+\frac13\,\frac{p_5t}{1+(p_5t)^2},\nonumber\\
&&a_{3\phantom{,5}}=
\frac1{24}\Big(\frac{p_1t}{1+(p_1t)^2}+\frac{p_2t}{1+(p_2t)^2}\Big)
-\frac16\,\frac{p_5t}{1+(p_5t)^2},\\
&&p_{1,2}=2\pm\sqrt{3},\quad p_{3,4}=\sqrt{2}\pm1,\quad p_5=1,
\qquad (p_1p_2=p_3p_4=1).
\label{radii}\end{eqnarray}
The remaining statements of the lemma follow from these expansions.
\endproof

We can now prove the following two theorems for magnetic field $B=0$:
\begin{theor}
The correlation functions $\langle\prod_{i=1}^m\sigma_{j_i}\rangle_N$
and their thermodynamic\b limits $\langle\prod_{i=1}^m\sigma_{j_i}\rangle$
are analytic, having series expansions in $t$ or $\beta J$
with radius of convergence bounded below by (\ref{suzidn}) and uniformly
convergent for all $N$ including $N=\infty$. Let $d$ be the largest
edge of the minimal parallelepiped containing all sites $j_1,\ldots,j_m$.
Then the coefficient of $t^k$ with $k<n-d$ for the lattice with $N=n^3$
sites equals the corresponding coefficient for larger $N$, including
the one for $N=\infty$.
\label{corrtheorem}\end{theor}

\Proof
We can assume that $m>0$ and even, since for $m$ odd we have
$\langle\prod_{i=1}^m\sigma_{j_i}\rangle_N$ $\equiv0$ as it both is invariant
and changes sign under the spin inversion $\sigma_i\to-\sigma_i$
for all sites $i$.

The system of equations (\ref{suzid})--(\ref{suzida}) can be viewed as
a linear operator on the vector space of linear combinations of all
correlation functions of the 3-dimensional Ising model. It is easy to
estimate the norm of this operator. Using the alternating sign property
of the $a_i$'s, it is easy to verify that $a_1$, $a_3$, and $a_5$ can
all be written as $t$ times a series in $t^2$, which three series
consist of positive terms only when $t$ is imaginary. This means that
each $|a_i|$ is maximal for given $|t|$ when $t$ is imaginary and
within the radius of convergence, i.e.\ $p_2$ in (\ref{radii}).

{}From the $32m$ terms in the right-hand side (RHS) of (\ref{suzid})
after applying (\ref{suzidt}), it follows then that we only need to study
\begin{equation}
6a_1+20a_3+6a_5=\frac{2t(t^2+3)(3t^2+1)}{(1+t^2)(1+14t^2+t^4)}
\label{suzidi}\end{equation}
for purely imaginary $t$ to find the desired upper bound $r$ for the
norm. Setting $t=\mathrm{i}x$ with $0<x<2-\sqrt{3}$ to stay within
the first pole of (\ref{suzidi}), we next define
\begin{equation}
r=\frac{2x(3-x^2)(1-3x^2)}{(1-x^2)(1-14x^2+x^4)},
\qquad\mbox{for }x=|t|.
\label{suzidm}\end{equation}
We then have that the RHS of (\ref{suzid}) is bounded by $rM$,
where $M=\max|\langle\sigma\cdots\sigma\rangle|$ with the maximum taken
over all $32n$ pair correlations in the RHS. (Obviously, $M\le1$ if
$\beta\ge0$ and real, but we shall not use this.) 
We can easily show that $r<1$ for
\begin{eqnarray}
&&|t|<(\sqrt{3}-\sqrt{2})(\sqrt{2}-1)=0.131652497\cdots,
\quad\hbox{or}\nonumber\\
&&|\beta J|<\arctan[(\sqrt{3}-\sqrt{2})(\sqrt{2}-1)]
=0.130899693\cdots.
\label{suzidn}\end{eqnarray}

To prove analyticity of $\langle\prod_{i=1}^m\sigma_{j_i}\rangle_N$
as a function of $\beta$ at $\beta=0$, we apply (\ref{suzid}) to it.
Then we apply (\ref{suzid}) to each of the $32m$ new correlations,
and we keep repeating this process ad infinitum. Since $\sigma_i^2=1$,
we will from some point on regularly encounter the correlation
with $m=0$, i.e.\ zero $\sigma$ factors, for which $\langle1\rangle=1$,
so that the iteration process ends there. Each other correlation
(with $m>0$) vanishes with at least one power of $t$, as can be seen
comparing e.g.\ (\ref{suzid}) and (\ref{suzida}). We conclude that
the iteration process generates the high-temperature power series
in $t$ to higher and higher orders, for arbitrary given size $N$
of the system.

To get the partial sum of the series to a given order, we only need
to keep the contributions for which the iteration process has ended
and expand all occurring $a_i$ as series in $t$. The sum of the
absolute values of the terms is bounded by $\sum r^j<\infty$ when
(\ref{suzidn}) holds. However, the original correlation function
is meromorphic with a finite number of poles away from the real $t$
axis for any finite $N$. Thus for sufficiently high order of
series expansion in $t$, the remainder term is arbitrarily small.
The only possible conclusion is that we have proved convergence of the
series expansion of $\langle\prod_{i=1}^m\sigma_{j_i}\rangle_N$ in
powers of $t$, uniform in $N$ with a finite radius of convergence
in the complex $t$ and $\beta$ planes bounded below by (\ref{suzidn}).

To prove the final statement of the theorem for finite $N$, we notice that
the above iteration process generates new correlations with the range of
the positions $j$ of the spins extended by one in a given direction.
As long as we do not go around a cycle (periodic boundary condition)
of the 3-torus, we do not notice any $N$-dependence. It takes at least
$n-d$ iteration steps to notice the finite size of the lattice.

Combining the convergence uniform in $N$ with the fact that more and more
coefficients converge with increasing $N$, we conclude that
$\langle\prod_{i=1}^m\sigma_{j_i}\rangle_N$ converges to a unique limit
as $N\to\infty$ for $|t|<2-\sqrt{3}$, with the properties stated
in the theorem.
\endproof

\subsection{The reduced free energy and its thermodynamic limit}

\begin{theor}
The reduced free energy $\beta f_N$ for arbitrary $N$ and its
thermodynamic limit $\beta f$ are analytic in $\beta J$ for sufficiently
high temperatures. They have series\b expansions in $t$ or $\beta J$
with radius of convergence bounded below by (\ref{suzidn}) and\b uniformly
convergent for all $N$ including $N=\infty$. The first $n-1$ coefficients
of these series for $N=n^3$ equal their limiting values for $N=\infty$.
\label{FreeEn}\end{theor}

\Proof
To prove analyticity of $\beta f$ in terms of $\beta$ at $\beta=0$ it
suffices to study the internal energy per site or the nearest-neighbor
pair correlation function, as
\begin{equation}
u_N=\frac{1}{N}\langle\mathcal{H}_N\rangle_N^{\phantom{y}}=
\frac{\partial(\beta f_N)}{\partial\beta}=
-3J\langle\sigma_{0,0,0}\sigma_{1,0,0}\rangle_N^{\phantom{y}},
\label{internal}\end{equation}
as follows from (\ref{Z_N}) and (\ref{f_N}).
Here $\sigma_{0,0,0}$ and $\sigma_{1,0,0}$ can be any other pair of
neighboring spins. The proof then follows from Theorem \ref{corrtheorem}
and integrating the series for $u_N$, using $Z_N|_{\beta=0}=2^N$,
implying $\lim_{\beta\to0}\beta f_N=-\log2$.
\endproof

\begin{remark}\rm
Adding a small magnetic field $B$ and generalizing the
steps in the above, we can conclude that all correlation functions
are finite for small enough $|\beta|$ and $|\beta H|$, so that there are no
Yang--Lee zeros \cite{YL,LY} near the $H=0$ axis for small $\beta$
and $H$. The proof can also be generalized to the case that the
interactions are anisotropic, i.e.\ $J,J',J''$ as in \cite{zdz}.
Then $Z_N$ is a Laurent polynomial in each of $\mathrm{e}^{\beta J}$,
$\mathrm{e}^{\beta J'}$, and $\mathrm{e}^{\beta J''}$, etc.
\end{remark}

\begin{remark}\rm
Similar results can be derived for the
low-temperature series, for\b example after applying the Kramers--Wannier
duality transform to the high-temper\-ature regime of the dual system
with spins in the centers of the original cubes and with
four-spin interactions around all cube faces perpendicular to the edges
of the original lattice \cite{BDI}.
\end{remark}

\begin{remark}\rm
It is straightforward to calculate the first few high-temperature\b
series coefficients of the free energy by the method described
in this section, with or without using the averaging in (\ref{suzid}).
They agree with the long series reported in \cite{AF} and earlier
works cited there. Zhang's free energy formula claimed for all finite\b
temperatures \cite{zdz} does not agree, as already the coefficients
of $\kappa^2\equiv t^2$ in (A12) and (A13) of \cite{zdz} differ. Zhang's
excuse that there are two expansions, one for finite $\beta$ and one
for infinitesimal $\beta$, violates general theorems, that apply to
more general models than the Ising model \cite{wmfc2,jhhp}. Here
this excuse is invalidated in detail by Theorem \ref{FreeEn}.
\end{remark}

\begin{remark}\rm
Zhang's spontaneous magnetization series is obviously wrong. In three
dimensions one should have $I-1=\mathrm{O}(x^6)$, with
$x\equiv\mathrm{e}^{-\beta J}$ in the low-temperature limit,
$x\to0$ ($J>0$), as each spin has six nearest neighbors \cite{jhhp},
rather than eight, which would result in the four-dimensional
$I-1=\mathrm{O}(x^8)$ presented by Zhang in (103) of \cite{zdz}.
\end{remark}

\begin{remark}\rm
The finite radius of convergence of the series expansions about $\beta=0$
is also hinted at by the fact that the zeros of $Z_N$ for $T=\infty$
occur for $B=\pm\mathrm{i}\infty$, $\beta B=\pm\mathrm{i}\pi/2$ \cite{jhhp2}.
For fixed temperature $T$ or $\beta=1/kT$ and $\beta J>0$ real the zeros
of $Z_N$ lie on the unit circle in the complex $\mathrm{e}^{-2\beta B}$
plane \cite{LY}, all located at $-1$ at infinite temperature \cite{jhhp2}
and spreading out with decreasing temperature until the zeros ``pinch'' $+1$
on both sides of the unit circle at and below the critical temperature,
in agreement with the theory of Yang and Lee \cite{YL,LY}. Zhang's claim
that this pinching at $+1$ also occurs at $\beta=0$ \cite{zdz,zdz2,zdz3,mz4}
is disproved by Theorem \ref{FreeEn}.
\label{zerorem}\end{remark}

\vglue-1pt
\begin{corollary}
As pointed out already in \cite{wmfc,wmfc2,jhhp}, all final
results of \cite{zdz} are proven wrong, as they do not agree within
a finite radius of convergence with the well-known series expansion
coefficients. This also means that the conjectures of \cite{zdz}
are falsified.
\end{corollary}
\vfill

\section{Further remarks and objections}

\subsection{Two series expansions for the same object}

In appendix A of \cite{zdz} Zhang claims to reproduce the
first 22 terms of the high-temperature series for the free
energy. But this is no more than reverse engineering,
fitting the known coefficients \cite{AF} to an integral transform
(A.1) or (74) in \cite{zdz} giving the first few
coefficients of the weight functions as given in (A.2).
There is no more information than the series results
provided by others, so that this does not constitute a
new result, as explained in \cite{wmfc,jhhp}.

As this construction this way is based on a conjectured integral
transform of weight functions that can only be reconstructed
from a few known series coefficients, it cannot be considered
an exact solution. Knowing this, Zhang conjectures
ad hoc above (A.3) on page 5400 another choice for the
weight functions, namely $w_x=1$, $w_y=w_z=0$, leading to
another high-temperature series for non-infinitesimal
temperatures, in violation of the rigorous result on the
uniqueness of the series expansion presented in
section 2. This is not sound mathematics \cite{wmfc}.

\subsection{Citations by other authors}

The outcomes of \cite{zdz} have been criticized in \cite{fb,ggkr},
as they disagree with recent high-precision Monte Carlo calculations
presented there. Both the position of the critical point and the
values of the critical exponents differ from the ones in \cite{zdz},
while the results of \cite{fb,ggkr} agree with those of many
others obtained by a variety of methods \cite{PV}.

One paper on a decorated three-dimensional Ising model \cite{sdc},
mapping this model exactly to the Ising model on a cubic lattice,
used Zhang's free energy \cite{zdz} as an approximate result in the
analysis. The experimental paper \cite{whmbt} states that their
result for the critical exponent $\Delta=2.0\pm0.5$ is consistent
with \cite{zdz}. However, the reported error bar is so large that
this means nothing. Moreover, paper \cite{bl} on the Heisenberg model
only briefly cites \cite{zdz} as an Ising reference.

The authors of \cite{lmnk,law,lmn} learned from \cite{zdz} the quaternion
setup of the transfer matrix of the three-dimensional Ising model,
which was well-known earlier, see e.g.\ \cite{Maddox,Camp1,Camp2}.%
\footnote{Maddox only published his final formula for
the free energy \cite{Maddox}. The details were discussed in a special
session, where also his error (the same as Zhang's first
error \cite{jhhp}) was discovered.} In Zhang's work \cite{zdz} this
is treated before the first error occurs with the Jordan--Wigner
transformation to Clifford algebra operators. His $P$'s and $Q$'s
do not anticommute \cite{jhhp,zdz3}.

\subsection{Advertising wrong critical exponents}
Klein and March took the exact Ising critical exponents
for dimensions $d=1,2,4$ together with the proposal of
\cite{zdz} for $d=3$ and made an ad hoc fit \cite{km} for
all real $1\le d\le4$. However, they failed to compare with the
results from $\varepsilon$-expansion \cite{epsilon,PV}, where
$\varepsilon=4-d$. This is a serious shortcoming, as the \cite{km}
formulae disagree with the $\varepsilon$-expansion exponents for
small $\varepsilon$ and fail the one foremost explicit test available.
It may also be noted that the extrapolated Ising exponents for $d=3$
from $\varepsilon$-expansion agree with those extracted
from series expansions and Monte Carlo calculations \cite{PV},
while differing from those presented in \cite{zdz}.

March and Zhang have followed this paper \cite{km} up with thirteen
publications, thus perpetuating the errors of the original work
\cite{zdz}. Some of these works compare Zhang's critical exponents
with those from experiments on CrBr$_3$ and Ni \cite{mz1,mz2}.
Nickel is known to have Heisenberg exchange interactions
and its critical exponent $\beta$ is about the accepted value
for the three-dimensional Heisenberg model, which is also
about Zhang's value wrongly claimed for Ising.

Comparing experiments with models needs a discussion of
the interactions in the experimental compounds, whether
Ising or Heisenberg, isotropic or anisotropic, short-range
or long-range, etc. No such analysis was presented. The
same objection can be brought up about section 2 of \cite{mz8}.

In \cite{mz3} critical exponents for the two- and
three-dimensional $q$-state Potts model are discussed.
Those for $d=2$ are by now well established, but the values
presented for $d=3$ cannot all be correct, as for the Ising
case $q=2$ the exponents of \cite{zdz} have been used.

In \cite{mz6,mz7} a new formula for critical exponent $\delta$
is given, improving the one in \cite{km}. The same objection
still applies, as again no comparison with $\varepsilon$-expansion
is made.

It is implied by the theory of Yang and Lee \cite{YL,LY}, that the
best experimental results on Ising exponents are to be expected
from measurements on liquid-gas transitions in simple substances.
March and Zhang have admitted that the exponents of \cite{zdz}
fail this test, see section~3 of \cite{mz8}. Their suggestion
that the experiment needs to be redone carries no credibility,
as the critical exponents measured in a number of similar
experiments are indeed typical
for Ising, see section 3.2.2 of \cite{PV}.

\subsection{Singularity of free energy at \boldmath$T=\infty$}

Several statements in section 5 of \cite{mz4} repeat and expand
on statements in \cite{zdz,zdz2,zdz3} contradicting rigorous theorems
discussed in section 2 above. For example, while it is correct that
the free energy $f$ diverges at $T=\infty$, this does not correspond
to a physical singularity, as the combination $\beta f$ is to be used.
Indeed, $\mathrm{e}^{-\beta f}$ relates to the normalization
of the Gibbs ensemble probability distribution and $\beta f$ is
the principal object of Theorem \ref{FreeEn}. Multiplying
$\beta f$ with $kT$ results in $f$ having a convergent Laurent
expansion with a leading pole at $T=\infty$ that has no physical
significance. Another point is discussed in Remark \ref{zerorem}
in section 2.

\subsection{False argument for \boldmath$\alpha=0$}

Paper \cite{mz5} addresses tricritical behavior. The authors
claim that the logarithmic divergence of the specific heat,
$\alpha=0$ (log) at tricritical points in three dimensions,
supports the similar value reported in \cite{zdz} for the Ising
critical behavior. However, this reasoning is flawed lacking
any theoretical basis and contradicts the accepted value
$\alpha=0.110\pm0.001$, see eq.~(3.2) and tables 3--7 of \cite{PV}.

\subsection{The \boldmath$\epsilon=d-2$ expansion}

In papers \cite{mz9,mz10} on Anderson localization the authors
say that $\epsilon=d-2$ is not a small parameter for $d=3$,
just like $\varepsilon=4-d$ of the $\varepsilon$-expansion is not.
This ignores that the best $\varepsilon$-expansion extrapolation
results agree remarkably well with those from series, Monte Carlo,
and experiment \cite{PV}. This argument to support \cite{zdz}
is again not valid.

\subsection{Higher dimensions}

The combinatorial sums defining the 3-dimensional Ising
model involve commuting spin variables and an interaction
energy that is a function of these spins. There is no reason
to introduce time and quantum mechanics in this classical
system, as is done in \cite{mz11}. On the other hand,
introducing the transfer matrix changes one space coordinate
to (discrete) imaginary time. After ``Wick rotation'' to real
time the 3-dimensional Ising model relates to a (2+1)-dimensional
quantum system. The fourth dimension introduced in \cite{zdz}
has only been used to obtain wrong results violating
rigorous results.

\subsection{Fractal dimensions based on wrong results}

In \cite{mz12} Zhang and March write down some proposals for
fractal dimensions. However, the values given for dimension 3
are based on incorrect results of \cite{zdz}.

\subsection{Unfounded Virasoro algebra}

The most recent paper \cite{mz13} uses the weight factors
of \cite{zdz} to introduce a Virasoro algebra in 3+1
dimensions. This is ad hoc and the notations in equation (5)
and seven lines below (6) there are not mathematically sound.
To take the real part of the absolute value of a phase
factor instead of just writing 1 makes no sense. Also, the
Virasoro algebra relates to an infinite dimensional
symmetry, which is only consistent with conformal
symmetry in two dimensions, see e.g.\ \cite{BP}. Therefore,
\cite{mz13} has fundamental errors.

\subsection*{Acknowledgments}

The author thanks Professor M.~E.~Fisher for a number of useful suggestions.
The research was supported in part by NSF grant PHY 07-58139.

\vspace{5mm}

\noindent {Department of Physics, Oklahoma State University\\
145 Physical Sciences, Stillwater, OK 74078-3072, USA}


\setcounter{section}{0}

\author{Jacques H.H. Perk}
\address{Department of Physics, Oklahoma State University,\\
145 Physical Sciences, Stillwater, OK 74078-3072, USA\\
E-mail: perk@okstate.edu}

\newpage

\def\b{\break}

\noindent {\it Jacques H.H. Perk}

\vspace{0.5cm}

\noindent {\bf  Erroneous solution of three-dimensional (3D) simple
orthorhombic Ising lattices II. Comment to the Response to
``Erroneous solution of three-dimensional (3D) simple
orthorhombic Ising lattices'' by Z.-D.\ Zhang}

\vspace{0.4cm}

\noindent {\ebf Summary}

{\small The response by Zhang and March to a recent comment on several
of their papers only adds further errors and misleading statements.}


\section{The Series Test is Decisive}

In their Response \cite{zdz0} to my Comment \cite{jhhp0} Zhang and March do not
really address the new criticism raised to their work. Contrary to what is said,
Comment \cite{jhhp0} contains very little that is in my earlier Comment
\cite{2jhhp} and Rejoinder \cite{2jhhp2}. There is some material from the
unpublished additions to the arXiv version of the Rejoinder, but that part is
much improved with several new details added in \cite{jhhp0}. There
are also some pages discussing papers published later. Therefore, the
statement ``hard to find anything new'' is wrong. Also, several statements
are not even addressed in the Response and cannot be covered with the
``unnecessary to repeat all of Zhang's responses''.

Zhang seems to demand that I only comment on the ``validity of the topologic
approach developed'', even though this is not precisely defined in any of his
papers, apart from the formulation of his two conjectures in \cite{2zdz}.
However, these conjectures 1 and 2 are not backed up by any quantitative
evidence in the original 117 page work. Their validity can at this moment
only be judged by the resulting free energy.

In section 3.1 of \cite{jhhp0}, I noted that Zhang expresses the free energy
by an integral transform, given in (49) in \cite{2zdz}, on unknown weight
functions $w_x,w_y,w_z$, without a clear convincing argument how
to get these weight functions. This is brought as a consequence of conjectures
1 and 2 \cite{2zdz} and it is analogous to saying that the free energy is a Fourier transform
of some unknown function, by itself an empty statement.

Zhang made two choices in \cite{2zdz}. The first one is fitting series (A2)
in \cite{2zdz} to the free-energy high-temperature series to as many terms
as are known in the literature; the other is choosing weights (1,0,0).
The first way gives no exact result, as one has no more than the known
series terms. The second way leads to a different series, with the first
nontrivial term differing from the known series; it is disproved by
the first few terms of the well-known high-temperature series, since these
have been rigorously established, also by the construction in \cite{jhhp0}.

The older proofs cited in \cite{2wmfc,2jhhp} are correct but not easy to read.
Therefore, I gave a much simpler proof with mathematical
precision in \cite{jhhp0}. My proof does not depend on the papers by
Lebowitz and Penrose \cite{2lp} and by Gallavotti et al.\ \cite{2gmr}, contrary
to what Zhang seems to suggest.

\section{Arguments for Phase Transition at \boldmath{$T=\infty$} Are Invalid}

Statements made in \cite{zdz0} about \cite{2lp,2gmr} are taken out of context.
The inequality $\rm Re \beta > 0$ on page 102 in \cite{2lp} is
needed when the gas model has no hard core. Section II, however,
opens with the statement that analyticity at $\beta = 0$ can be shown
for a hard core potential. The Ising model is equivalent to a lattice
gas version with at most one particle per lattice site (empty-occupied
becomes spin~$+/-1$), a special case of a hard core on the lattice.
Thus the objection that \cite{2lp} excludes $\beta = 0$ in their
analyticity proof does not apply.

One statement in \cite{zdz0} about an inequality in \cite{GMS} not
being valid for $\beta=0$ is misplaced for two reasons. First, the
inequality does not appear in \cite{GMS}, but appears near the
bottom of the left column of page 494 of \cite{2gmr}. Secondly,
in order to prove a finite radius of convergence one needs to
prove an inequality with some positive $\beta$.
Then $\beta = 0$ will be included within the radius of convergence.
(It may be noted that there are misprints in \cite{2gmr}, probably due
to printer errors as Phys.\ Lett.\ did not allow
authors to correct proofs at that time.)

The next objection in \cite{zdz0} that $f$ is singular at $T = \infty$ is
also misleading. The combination $\beta f = -\ln 2$ there, as $f$ has a
simple pole at $\beta =1/k_{\mathrm{B}}T= 0$. One finds that $f$ has
a Laurent expansion
with pole term $(-\ln 2)/\beta$ followed by a power series in $\beta$
with a finite radius of convergence. Statistically, at $\beta = 0$ all
states have equal probability and there is no phase transition, as
not only $\beta f$, but also all correlation functions are analytic
at $\beta=0$, in spite of the fact that interactions are turned on
once $\beta>0$.

When Zhang expands $\lambda=Z^{1/N}$ in (A12) and (A13) of \cite{2zdz},
he expands $\exp(-\beta f)$, which is equivalent to expanding $\beta f$.
This makes his objection to expanding $\beta f$ instead of $f$
unreasonable. The finite radius of convergence proof given in \cite{jhhp0}
proves that (A13) is not correct.

Response \cite{zdz0} bring up that $1/Z$ has a zero at $\beta = 0$ in the infinite
system. But this is again misleading. Yang--Lee theory is only about the
zeroes of partition function $Z$: When zeros pinch the real temperature
axis in the large system limit, then there is a phase transition. There
is no theorem for $1/Z$.

This pinching of zeros of $Z$ cannot occur, as the proof given
in \cite{jhhp0} can be extended to
the double expansion of $\beta f$ in $\beta$ and $\beta B$. The proof
for the more general cases is in the old literature. From this joint
analyticity at $\beta = 0$  and $\beta B = 0$, it follows that zeros
are a finite (nonzero) distance away, contradicting any pinching of
zeros at $\beta =\beta B = 0$.

\section{Other Issues}

Remark 2.4 in \cite{jhhp0} and similar statements show that it is possible to
test Zhang's free energy with Monte Carlo methods \cite{2fb}, as one can now estimate
both the systematic error due to finite size and the statistical error
due to Monte Carlo methods.

Also, the latest various experimental and theoretical estimates for
$\alpha$ are significantly different from 0, see the review by Pelissetto
and Vicari \cite{2PV}. One may want to check the accuracies
reported of large numbers of theoretical and experimental works that
are discussed there and ignored by March and Zhang.

Next, the use of dimensionless $K_i=\beta J_i$ and $h=\beta H$ can be
done in more than one way. The partition function and correlation
functions (and $\beta f$) only depend on these combinations. That some
authors set $\beta=1$, does not mean a loss of the high-temperature case.
If one has the result in the $K_i$ and $h$, one also can choose a
new $\beta$, say $\beta'$, and write in the results $K_i=\beta' J_i$
and $h=\beta' H$. There is no loss of the high-$T$ limit $\beta'=0$.
Again this is an objection that is invalid and it does not apply to
\cite{jhhp0}, as I nowhere used $\beta=1$, nor did I use results
from authors that did.

Finally, the last paper \cite{2mz13} is based on an incorrect
solution of the 3D Ising model. There are problems I noted: With
$\phi$ a phase, $|e^{i\phi}|=1$, and formula (4) contains phases
that drop out. Also, having three independent Virasoro algebras
means that one has the 3+1 dimensional space rewritten as a
6-dimensional (2+2+2) space, the ``product'' of three independent
2-dimensional spaces. Things do not add up.

\subsection*{Acknowledgments}

The research was supported in part by NSF grant PHY 07-58139.

\vspace{5mm}

\noindent {Department of Physics, Oklahoma State University\\
145 Physical Sciences, Stillwater, OK 74078-3072, USA}


\setcounter{section}{0}

\author{Jacques H.H. Perk}
\address{Department of Physics, Oklahoma State University,\\
145 Physical Sciences, Stillwater, OK 74078-3072, USA\\
E-mail: perk@okstate.edu}

\newpage

\def\b{\break}

\noindent {\it Jacques H.H. Perk}

\vspace{0.5cm}

\noindent {\bf  Erroneous solution of three-dimensional (3D) simple
orthorhombic Ising lattices III. Rejoinder to the 2nd Comment (Rejoinder)
to the Comment to the Response to ``Erroneous solution of three-dimensional
(3D) simple orthorhombic Ising lattices'' by Z.-D.\ Zhang}

\vspace{0.4cm}

\noindent {\ebf Summary}

{\small The responses by Zhang and March to recent comments on several
of their papers are questionable, misleading and outright wrong.}


\section{The series test is all-deciding}

In 2007 Zhang claimed to have an exact solution of the
three-dimensional Ising model \cite{3zdz}. However, Zhang's ``putative''
solution is wrong failing the series test \cite{3wmfc} and this has
been backed up recently by a detailed elementary proof \cite{3jhhp0}.
It is, therefore, surprising that Zhang and March \cite{3zdz0,3zdz1,3zdz2}
still defend the refuted claims of \cite{3zdz}, even after a subsequent
comment \cite{3jhhp1}. No further comments should be necessary, but
being accused of perpetrating fraud at the end of section 2 of \cite{3zdz0}
and in section 4.4 of \cite{3zdz2} I feel compelled to respond one more
time with further details.

The proof given in \cite{3jhhp0} shows that $\beta f$, with $f$
the free energy per site and $\beta=1/k_{\mathrm{B}}T$ inversely
proportional to the absolute temperature $T$, is analytic in $\beta$
with a finite radius of convergence $r$, i.e.
\begin{equation}
\beta f=\sum_{i=0}^{\infty}a_i\beta^i,\quad |\beta|<r.
\end{equation}
In Appendix A of \cite{3zdz} Zhang gave two different series expansions
for $\log\lambda=-\beta f$ in powers of $\tanh K$ with $K=\beta J$ and
from this it was already manifest that the ``putative'' solution has a
series expansion different from the well-known high-temperature expansion,
see also section 3.1 of \cite{3jhhp0} for more discussion.

To counter the series test failure, Zhang and March \cite{3zdz1,3zdz2}
claim that tests should be applied to $f$, which equals $-\infty$
at $\beta=0$. They claim that this singularity invalidates the application
of the series test at $\beta=0$. However, this argument is misleading as
\begin{equation}
f=\frac{a_0}{\beta}+\sum_{i=1}^{\infty}a_i\beta^{i-1}
\end{equation}
is a convergent Laurent series with the same coefficients, so that it has
to obey the same series test. The simple pole at $\beta=0$ is of no
consequence, as in statistical mechanics the partition function per site
$z$, $\log z=-\beta f$, is more fundamental. It relates to the
normalization $Z=z^N$ of the Boltzmann--Gibbs canonical ensemble
distribution of the Ising model with $N$ sites, in the large-$N$ limit.

The failure of the series test raises the question if one of the
conjectures in \cite{3zdz} fails and it is now easy to show that this
is the case.

\section{Conjecture 1 is manifestly wrong}

The original paper \cite{3zdz} has an error in the application
of the Jordan--Wigner transformation pointed out in \cite{3jhhp}.
This error has only been corrected explicitly in a very recent
paper \cite{3zdz2}, which makes it easy to pinpoint the error with
Conjecture 1 \cite{3zdz,3zdz2}.

It is well-known that in the spinor representation of the orthogonal
groups each element $g$ can be written as a ``fermionic Gaussian'' of
the form
\begin{equation}
g=\exp\bigg(\sum_i\sum_j A_{ij}\Gamma_i\Gamma_j\bigg),
\quad A_{ji}=-A_{ij},
\label{spinor}\end{equation}
with Clifford algebra elements satisfying
$\Gamma_i\Gamma_j+\Gamma_j\Gamma_i=2\delta_{ij}$ and complex coefficients
$A_{ij}$. In \cite{3Kaufman,3zdz} the $\Gamma_i$'s are also written as
$P_i$'s and $Q_i$'s and we note that the $\Gamma_i$'s can be expressed
as linear combinations of
fermion creation and annihilation operators $c_i^{\dagger}$ and
$c_i^{\vphantom{\dagger}}$. The spinor representation has been used in the
Ising context by Kaufman \cite{3Kaufman} in 1949. The closure property
of Lie groups says that any product or inverse of elements of the form
(\ref{spinor}) is again of the same fermionic Gaussian form.

The factors in (15) of \cite{3zdz2} commute, compare (7a) in \cite{3zdz},
so that $\mbox{\boldmath{$V$}}_3$ can be rewritten as
\begin{equation}
\mbox{\boldmath{$V$}}_3=\exp{\bigg(-\mathrm{i}K''\sum_{j=1}^{nl}
\Gamma_{2j}\bigg[\prod_{k=j+1}^{j+n-1}\mathrm{i}\Gamma_{2k-1}\Gamma_{2k}
\bigg]\Gamma_{2j+2n-1}\bigg)},
\end{equation}
which is clearly not of the form (\ref{spinor}) and, therefore, not an
element of the group. Multiplying $\mbox{\boldmath{$V$}}_3$ with group
elements $\mbox{\boldmath{$V$}}_2$, $\mbox{\boldmath{$V$}}_1$ and
$\mbox{\boldmath{$V$}}_4'$, as given in (16), (17) and (18) in \cite{3zdz2},
cannot give a group element as implicitly claimed in \cite{3zdz2}. Zhang
and March violate a fundamental property of Lie groups. Hence, Conjecture 1
as stated in \cite{3zdz2} fails.

\section{Zeros of $Z^{-1}$ are irrelevant}

Zhang and March repeatedly \cite{3zdz0,3zdz1,3zdz2} claim the importance
of the zeros of $Z^{-1}$, with $Z=z^N$ the total partition function.
However, unlike the complex Yang--Lee zeros of $Z$ \cite{3YL,3LY},
the zeros of $Z^{-1}$ are irrelevant. For a finite number $N$ of sites,
$Z$ is a finite Laurent polynomial in $\mathrm{e}^K$ and only can become
infinite when $\mathrm{Re} K=\pm\infty$, i.e.\ zero-temperature type limits.

For the infinite system, $N\to\infty$, and finite $K$, the infinity of $Z$
should be seen as just a manifestation of the thermodynamic limit, in which
$z=Z^{1/N}$ remains finite. One can easily see that $\beta f<0$ for $K$
real, so that $Z=\mathrm{e}^{-N\beta f}=\infty$ for all real $K$ when
$N=\infty$. There is nothing special about the zeros of $Z^{-1}$.
If Zhang and March want to claim differently they face inconsistencies
even in the one-dimensional Ising model.

\section{Other issues}

Bringing up \cite{3lp,3gmr} at great length is only a smoke screen, as
\cite{3jhhp0} made no use of these references and provides an independent
derivation. Zhang and March also misrepresent statements in section 3
of \cite{3jhhp1}: Setting $\beta=1$ in \cite{3lp,3gmr} is no loss of
generality, as $\beta f$ is only a function of $K=\beta J$. Having
$J\equiv K$ and choosing a fixed $\bar J$ and a new $\beta=J/\bar J\not\equiv1$,
we can write $J=K=\beta\bar J$. Thus we recover the general case with
both a fully variable $\beta$ (including $\beta=0$) and a new $J$ (omitting
the bar on $\bar J$). This is said in another equivalent way in section 3
of \cite{3jhhp1}.

Next, Zhang and March fail to realize that $K_{\beta\phi'}(X,T)$
vanishes for $\beta=0$, so that the inequality in \cite{3gmr} does not
fail for $\beta=0$, contrary to what is said in \cite{3zdz0,3zdz1,3zdz2}.
Also, this inequality plays no role in the proof of \cite{3jhhp0}, so that
bringing it up can only be seen as a diversion.

Finally, only in two dimensions is the conformal group infinite-dimensional,
so that there is inconsistency in \cite{3mz13,3zdz2} beyond the fact that
these papers build on an erroneous solution of the 3D Ising model. That
Zhang and March write $\mathrm{Re}|\mathrm{e}^{\mathrm{i}\phi_i}|$, the real
part of a positive real number, is objectionable too.

\section{Conclusion}

As should already have been clear from \cite{3wmfc,3jhhp}, Zhang's very
long paper \cite{3zdz} and all the works building on it are in error.
Some further errors have been shown explicitly above, including why
Conjecture 1 does not hold. 

\subsection*{Acknowledgments}

The research was supported in part by NSF grant PHY 07-58139.

\vspace{5mm}

\noindent {Department of Physics, Oklahoma State University\\
145 Physical Sciences, Stillwater, OK 74078-3072, USA}


\author{Jacques H.H. Perk}
\address{Department of Physics, Oklahoma State University,\\
145 Physical Sciences, Stillwater, OK 74078-3072, USA\\
E-mail: perk@okstate.edu}

\end{document}